*Review*

# Planetary Nebulae Research: Past, Present, and Future

Sun Kwok [1,2]

1 Department of Earth, Ocean, and Atmospheric Sciences, University of British Columbia, Vancouver, BC V6T 1Z4, Canada; sunkwok@hku.hk
2 Laboratory for Space Research, Faculty of Science, The University of Hong Kong, Hong Kong, China

**Abstract:** We review the evolution of our understanding of the planetary nebulae phenomenon and their place in the scheme of stellar evolution. The historical steps leading to our current understanding of central star evolution and nebular formation are discussed. Recent optical imaging, X-ray, ultraviolet, infrared, millimeter wave, radio observations have led to a much more complex picture of the structure of planetary nebulae. The optically bright regions have multiple shell structures (rims, shells, crowns, and haloes), which can be understood within the interacting winds framework. However, the physical mechanism responsible for bipolar and multipolar structures that emerged during the proto-planetary nebulae phase is yet to be identified. Our morphological classifications of planetary nebulae are hampered by the effects of sensitivity, orientation, and field-of-view coverage, and the fraction of bipolar or multipolar nebulae may be much higher than commonly assumed. The optically bright bipolar lobes may represent low-density, ionization-bounded cavities carved out of a neutral envelope by collimated fast winds. Planetary nebulae are sites of active synthesis of complex organic compounds, suggesting that planetary nebulae play a major role in the chemical enrichment of the Galaxy. Possible avenues of future advancement are discussed.

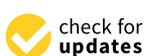

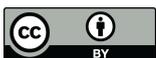

## 1. Introduction

The history of planetary nebulae research goes back over 200 years. Four objects (M27, M57, M76, M92) in Charles Messier's 1784 catalog of nebulous objects were later included in a list of objects called planetary nebulae (PNe) by William Hershel, because their appearance resembled the greenish disk of the planet, Uranus. In 1864, William Huggins took a spectrum of NGC 6543 and found emission lines, therefore concluding that it is a gaseous nebula, not a collection of stars. An observed correlation between the magnitude of the central star and the size of nebula led Edwin Hubble in 1922 to suggest that the nebula derives its energy from the central star [1]. After advances in atomic physics in the early 20th century, Donald Menzel theorized that the nebula is ionized by Lyman continuum photons from the central star [2]. Based on this assumption, Herman Zanstra was able to derive the temperature of the central star from the nebular Hβ flux and concluded that the central stars of PNe are much hotter than normal stars [3]. In 1928, Ira Bowen identified the nebular lines found by Huggins to be due to metastable states of $N^+$, $O^+$, and $O^{++}$ [4]. These lines do not originate from a new element nebulium, as previously believed, but are just ordinary matter radiating under very low-density conditions. These successful applications of atomic physics to astronomy represented the beginning of the discipline of astrophysics.

The inventory of PNe continued to increase through the discovery of new PNe from catalogs of extended nebular objects [5]. Objective prism surveys also reveal new emission-line objects and candidates of PNe [6,7], resulting in comprehensive catalogs of PNe [8,9]. Recent narrow-band surveys have produced a count of over 3800 PNe in the Galaxy [10].

From the 1920s on, research on PNe focused on atomic spectroscopy and nebular physics. This led to a better understanding of the process of photoionization and recombination of H, and the derivation of the chemical abundance of elements through observation of the recombination lines of H and He, and collisionally excited lines of metals (O, N, S, Ne, . . . ) [11].

In this paper, we discuss the evolution of our understanding of PNe in the context of stellar evolution, observational advances made possible by modern multi-wavelength observing techniques, and the challenges we face today in achieving a comprehensive understanding of the physical and chemical structures of PNe.

## 2. PN as a Phase of Stellar Evolution

In the early 20th century, stars were believed to evolve from high to low temperatures along the Main Sequence. Having hot central stars, PNe were thought to be young stars. However, the Galactic distribution of PNe suggests that PNe belong to the old star population [12]. The modern interpretation of PNe in the scheme of evolution was proposed by Josif Shklovsky, who suggested that PNe descend from red giants and are progenitors of white dwarfs [13]. Recognizing that the expansion velocities of PNe are similar to the escape velocities of red giants, Abell and Goldreich suggested that PNe are ejected envelopes of red giants [14]. To place the central stars of PNe on the Hertzsprung–Russell (H-R) diagram, their luminosities were derived from the apparent magnitudes of the central stars and distances determined by the Shklovsky method [15], together with the central star temperatures determined by the Zanstra method [16,17]. The result was the Harman–Seaton sequence, which suggested that PNe were formed at the end of the horizontal branch, followed by a rapid rise in temperature and luminosity [18]. Theoretical models of evolution were developed in an attempt to reproduce the Harman–Seaton sequence [19]. An editorial published in the September 30, 1967 issue of *Nature* states that "Theoretical models agree tolerably well with the observed data . . . the overall properties of planetary nebulae are now well understood" [20].

These predictions were overly optimistic. There were many efforts to create evolutionary tracks to simulate the Harman–Seaton sequence [21–24]. These models all had difficulty achieving a rapid change in temperature and luminosity within the time scale dictated by the dynamical age (~$10^4$ yr) of the nebula. It turned out that there were many problems with the observed H-R diagram for PNe. There was not enough accounting for the missing ultraviolet flux due to nebulae not being ionization bounded, and a lack of accounting for the infrared flux emitted by the dust component. The statistical distances based on the Shklovsky method also yielded unreliable distances [25], therefore they gave wrong luminosities. Due to the faintness of the central star in terms of visibility because of their high temperature (30,000–300,000 K) and contamination by nebular emission, the temperature measurements of the central star were often incorrect.

The inability of nuclear burning models to reproduce the Harman–Seaton sequence was finally resolved with a simple theory by Bohdan Paczynski [26]. Paczynski assumed that PNe descend from double-shell burning, asymptotic giant branch (AGB) stars. He started with a static stellar structure model of AGB stars, by putting a H envelope above an electron-degenerate carbon–oxygen core and observing the change in the effective temperature ($T_{\rm eff}$) of the star by reducing the mass in the H envelope. He found that the effective temperature of the star would not increase until the envelope mass was reduced below $10^{-3}$ $M_\odot$ for a core mass of 0.6 $M_\odot$. For higher core masses, the required shell mass is even lower. When the central star begins to increase in temperature, the H envelope mass is so low that the mass of the core is unchanged by nuclear burning. Because of the core mass–luminosity relationship, central stars of PNe evolve at a constant luminosity, in contrast to the Harman–Seaton sequence.

The evolution of the central star across the H-R diagram is the result of a combination of the effects of thinning of the H envelope due to H burning and mass loss. In Pacynski's model, a higher core-mass star will require a lower threshold envelope mass before evolving.

The combined effect of a lower remnant envelope mass and a higher burning rate for high-mass star causes the evolutionary time across the H-R diagram to be highly sensitive to the core mass ($M_c$):

$$t \propto M_c^{-8}$$

When Paczynski presented his results to the International Astronomical Union Symposium on PN in 1976, most of the participants were skeptical. His model gradually gained acceptance as people realized that the observational diagram was not reliable. The Paczynski models were extended by Schönberner [27,28], Kovetz and Harpaz [29], Iben [30], and Wood and Faulkner [31]. A recent update to the H-shell burning model of central stars of PNe is given by Miller Bertolami [32].

According to these H-shell burning models, the existence of PNe requires the dynamical age of the nebula to be comparable to the post-AGB evolutionary time of the central star. Since the transition time is highly dependent on the core mass, the lifetime of high-mass PNe are very short. Since low-mass PNe evolve very slowly, the nebulae will dissipate before the central stars reach a high enough temperature to photoionize the nebula. The upper limit of the PN central star mass is the Chandrasekhar limit (1.4 $M_\odot$) and the lower limit is set by the PN core mass, below which evolution is too slow to ionize the nebula before it disperses. Because of the initial mass function of stars, the mass distribution of PN central stars is confined to a narrow range, with a peak around $M_c \sim 0.6$ $M_\odot$ [33,34].

## 3. Formation of the Nebula

Since PNe consist of both a central star and a nebula, the next question to tackle after the evolution of the central star concerns the formation and dynamical evolution of the nebula. How is the nebula ejected and what is driving its expansion?

### 3.1. Ejection of the Hydrogen Envelope

After the work by Abell and Goldreich [14] showing that PNe are ejected envelopes of red giants, many theories were developed for the mechanism of ejection. The proposed mechanisms include dynamical instabilities [19,35,36], pulsational instabilities [37–39], envelope oscillations due to thermal instabilities in the core [40], radiation pressure [41–43], and thermal pulses [44,45]. Since sudden ejection is a one-time event, gravity from the central star may cause some of the ejected material to diffuse back towards the central star and additional pressure is needed to maintain the shell structure of the nebula [46–49].

All sudden-ejection theories faced an additional problem after the work of Paczynski showed that the central star will not evolve until the envelope mass falls below $10^{-3}$ $M_\odot$. No sudden ejection can be so precise as to leave behind such a small amount of envelope mass. For example, even if 0.199 $M_\odot$ of a 0.2 $M_\odot$ H envelope is ejected, the star will remain red, and the ejected envelope will not be photoionized before it disperses into the interstellar medium. No PN will be observed.

### 3.2. The Interacting Winds Theory

The development of infrared and millimeter wave astronomy in the late 1960s led to the discovery of mass loss by stars on the AGB. AGB stars are observed to be losing mass via stellar wind. The mass-loss rates of some AGB stars are so large that circumstellar dust completely obscures the photosphere of the central star. Estimates of the mass of the circumstellar envelope from CO observations have in some cases exceeded one solar mass, suggesting that a significant fraction of the total mass of the star can be ejected via a stellar wind. Since the nebular mass of PNe is typically only around 0.2 $M_\odot$, remnants of the circumstellar envelope ejected during the AGB phase would be expected to have an effect on the formation of PN.

Assuming that the nebular mass of PNe is provided by the AGB wind, the interacting stellar winds theory was developed to explain the formation of PNe. After the H envelope of an AGB star is completely depleted by a slow wind, the exposed hot and dense core will develop a much faster wind, which can compress the remnant AGB wind into a dense shell, and accelerate the shell to a typical PN expansion velocity of 25 km s$^{-1}$ from the typical 10 km s$^{-1}$ speed of the AGB wind [50,51]. Details of the interacting winds dynamics and the related predictions are summarized in reference [52].

Confirmation of the predicted fast wind from central stars of PNe came with the *International Ultraviolet Explorer (IUE)* observations of P Cygni profiles in central stars of PNe [53]. The remnants of AGB winds outside of the ionized PN shell were detected by molecular line observations. The *Infrared Astronomical Satellite (IRAS)* found large infrared excesses in PNe, confirming that remnants of AGB envelopes are still present in PNe [54,55]. The development of charged-coupled devices (CCDs) allowed high-dynamic range optical imaging and the detection of optical haloes outside the main shells of PNe [56–58]. Finally, the predicted hot bubbles from the shocked fast wind were detected by X-ray observations with the *ROSAT* and *Chandra* satellites [59–61].

The interacting winds theory also naturally complements Paczynski's theory on central star evolution. The H envelope is gradually removed by the stellar wind and the star will stay at a low (<5000 K) effective temperature until the critical envelope mass is reached. The continued thinning of the envelope by mass loss causes the star to gradually appear hotter and evolve to the left of the H-R diagram; therefore, avoiding the problem of uncertain remnant mass in sudden-ejection theories, mentioned in Section 3.1.

A PN is a dynamical system whose evolution is tightly coupled to the evolution of the central star. The changing luminosity and effective temperature of the central star leads to a changing mass loss rate and velocity of the central star's wind. The nebular shell is ionized by ultraviolet photons from the central star, and this photoionization rate is also a function of time. The coupled central star evolution and interacting winds dynamics produces a changing density, velocity, and thickness of the nebular shell [62–66]. Asymmetric morphologies of PNe could develop as the result of interacting winds dynamics, if the AGB wind is not spherically symmetric [67–71].

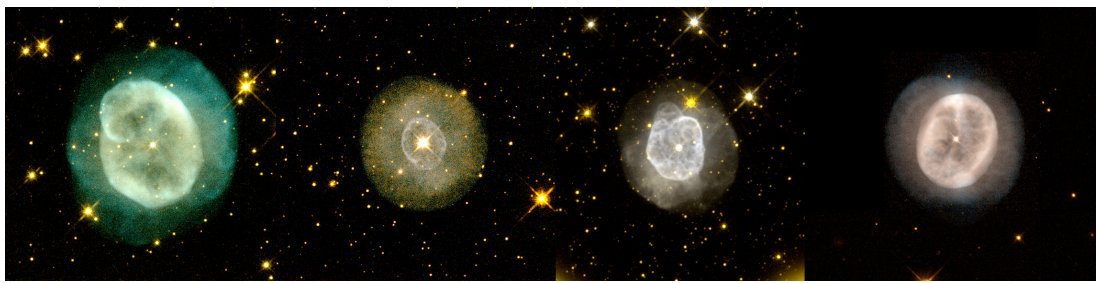

**Figure 1.** Examples of optical outer structures revealed by *HST* imaging. From left to right: NGC 5979, NGC 6629, NGC 6578, NGC 2022. Images from reference [72]. See references [58,73,74] for descriptions of outer structures of PNe.

Even for spherically symmetrical PNe, their structures are not simple. PNe often show multicomponent structures, including rims, shells, crowns, and haloes [73–75] (Figure 1). Such multiple shells can only be understood with the consistent treatment of dynamics and stellar evolution [76–78].

## 4. Missing Link between AGB and PN

The evolutionary stage between the end of the AGB and the beginning of the PN phases had long been a missing link in our understanding of single star evolution. The transition phase between the end of the AGB and the photoionization of the nebula is called the proto-planetary nebula phase. Proto-planetary nebulae (PPNe) are difficult to detect because they do not have emission lines. The only source of illumination of the nebula is from scattered starlight, which can be very faint. Because they are young, the nebula is often small, and resembles ordinary stars in optical images.

After the *IRAS* mission, it was found that late AGB stars and young PNe have distinct infrared colors, and candidates for PPNe can be found by searching for objects with *IRAS* intermediate colors intermediate between the two groups [79,80]. This new class of PPNe was discovered through the ground-based identification of *IRAS* sources and follow-up photometric and spectroscopic observations [81,82]. Some of these PPN candidates were imaged by the *Hubble Space Telescope* (*HST*). Figure 2 shows the images of PPNe IRAS17150–3224 [83], IRAS17441–2411 [84], IRAS16594–4656, and IRAS 17245–3951 [85]. These images clearly show that the shaping of PN morphology has already started in the PPN phase.

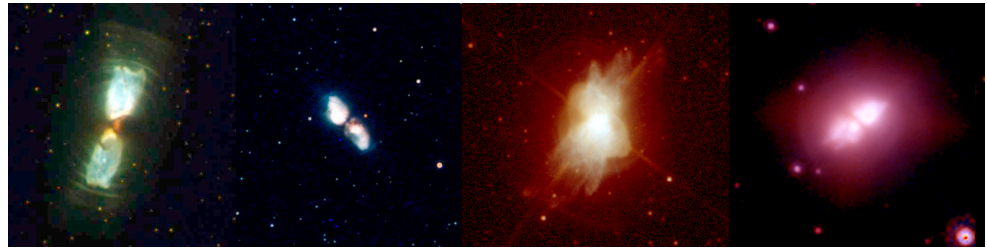

**Figure 2.** *HST* images of 4 proto-planetary nebulae. From left to right: IRAS 17150–3224 (Cotton Candy Nebula), IRAS17441–2411 (Silkworm Nebula), IRAS16594–4656 (Water Lily Nebula), and 17245–3951 (Walnut Nebula). The common names are given by Sun Kwok and Bruce Hrivnak, who discovered these nebulae. Pictures adapted from reference [72].

Candidates of stars in the post-AGB phase of evolution have been identified using a variety of criteria [86]. *HST* images of these candidates show a variety of degrees of asymmetry [87–89]. The difference in morphology can in part be explained by orientation effects [90].

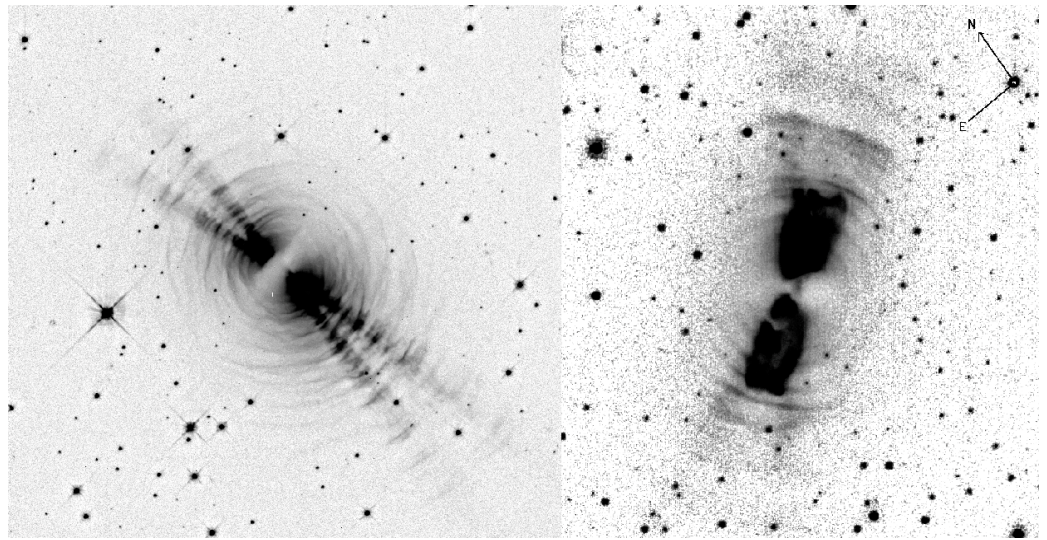

**Figure 3.** *HST* 606W image of AFGL 2688 (**left**) and IRAS 17150–3224 (**right**), showing a series of concentric circular arcs. Assuming distances of 1 kpc for AFGL 2688 and 2 kpc for IRAS 17150–3224, the time intervals between the arcs are 160 and 540 yr, respectively. These arcs are possible manifestations of episodic mass loss during the AGB phase. Figure adapted from reference [91].

An interesting morphological feature of PPNe is the presence of multiple concentric arcs. They were first seen in the Egg Nebula AFGL 2688 [92]. A series of concentric arcs, separated by nearly equal intervals, have been found in many PPNe [83,84,91,93,94] (Figure 3). These arcs are probably 3D thin shells projected onto the sky and are different from the multiple 2D rings that are observed in PNe [95,96]. The possible origin of arcs and rings, as a result of a bi-conal cavity carved out by fast outflows in a multiple spherical shells, is discussed in terms of a 3D model [97].

## 5. Binary Evolution

So far, we have described the formation of PNe as arising from an evolving single star. But many stars have at least one stellar companion. How might this influence its evolution? For example, a companion star may gravitationally alter the shape of the AGB wind, or the expansion of the envelope of the AGB star may engulf the companion, resulting in a common envelope situation. These possibilities may cause changes in the morphology or even the evolutionary path of PNe.

Our understanding of binary star evolution has undergone major developments since the 1960s. The identification of novae as the result of binary evolution [98] represented the beginning of our understanding of cataclysmic variables (CVs) [99]. We now realize that a binary system can undergo mass transfer (through Roche lobe or stellar wind accretion) when either the primary or the secondary component is in the post-main-sequence phase of evolution.

Binary star evolution can involve a large number of evolutionary scenarios, depending on the initial mass of the binary components and the initial binary separations [100]. Depending on the respective stellar evolutionary stages of the two components when mass transfer occurs, a flow chart of binary evolution suggests that over one hundred different types of binaries can be products of evolution [101,102].

Like PNe, classical novae also have gaseous envelopes and a hot central star. As the result of accretion, novae can reignite nuclear burning on the surface of the white dwarf component. The subsequent ejection of envelope material can lead to an increase in the size of the pseudo-photosphere and a decrease in the effective temperature of the central star. This lowering of the effective temperature leads to an increase in brightness in optical wavelengths, while the total luminosity of novae remain the same [103]. In other words, classical novae are white dwarfs evolving backwards on the H-R diagram. From an evolutionary point of view, novae are very different objects from PNe.

Close binary systems can form common envelopes [104–108], which can lead to common envelope ejections [109]. In recent years, there has been increasing interest in the idea that PN is a CV phenomenon, as some PNe are found to have binary central stars [110–112]. If PNe are the result of binary evolution, then we must address the question: what kind of CV is it? How did it evolve to this stage, and how will it evolve in the future? What happens to the central star after common envelope ejection? Will it evolve? If not, do we still call it a PN? Where do PNe fit into the many branches of the binary evolution chart?

Even if a PN can be produced through a CV-like evolution, this would not imply that all PNe are formed that way. We also should remember that before we have a CV, we first need to have a white dwarf. The existence of a white dwarf in a CV implies that there must be a way for a red giant to evolve to a white dwarf via single star evolution. The PN phenomenon can be part of this single star evolution scenario.

## 6. Observational and Evolutionary Definition of PN

Central to every topic of scientific enquiry is the classification and definition of the objects under study. Unfortunately, there is a lot of confusion in the case of PNe. In the beginning, PNe were defined by their morphology: a circumstellar nebula showing some degree of symmetry. With the advent of spectroscopy, PNe were defined as circumstellar nebular objects showing an emission-line spectrum with little or no continuum. Although PNe share some common observational properties with other emission-line nebulae, such as H II regions, they can be distinguished by their spectral properties. Compared to H II regions, PNe have lower H$\alpha$ to [O III] ratios. They also show characteristic expansion velocities of 20–50 km s$^{-1}$. The central stars of PNe have high temperatures (>30,000 K), high luminosity (>$10^3$ L$_\odot$), and high surface gravity (log g~4.5–7) [113]. Additional observational properties, such as surface brightness, have also been proposed [114].

Objects that are frequently misidentified as PNe are emission-line galaxies, reflection nebulae, H II regions, Wolf–Rayet nebulae, novae, supernovae remnants, young stellar objects, and symbiotic stars. The most common source of confusion is with symbiotic stars. Symbiotic stars are defined as objects showing the simultaneous presence of spectral characteristics of cool stars and nebular emission spectra as the result of photoionization by a companion hot star [115]. They, therefore, have similar observational nebular properties to PNe.

Symbiotic stars are binary systems consisting of a cool red giant (on the red giant branch or AGB) and a white dwarf [116]. One way to distinguish a symbiotic star from a PN is to search for photospheric signatures of the red giant. If an object shows $H_2O$ absorption bands and photometric variability, it is most likely a symbiotic star [117]. Evolution in a symbiotic binary system is driven by the mass transfer rate and H burning rate [118]. In the binary star evolution scenario, symbiotic stars are static, not evolving backwards as in the case of classical novae or evolving forward as in the case of PNe.

Since an observational definition cannot uniquely define a PN, we may need to adopt a definition that includes its status in stellar evolution. For example, we can define a PN as an emission-line object showing an ionized circumstellar shell with some degree of symmetry surrounding a hot, compact star evolving from the AGB to the white dwarf stage. This combined theoretical–observational definition will put PNe in a unique class of objects in the scheme of stellar evolution.

## 7. Morphological Classifications

Although PNe are commonly perceived to have spherical shells, they in fact have a rich variety of morphologies. In the early 20th century, Curtis classified the PNe he observed with the Crossley reflector at Mount Hamilton into classes of helical, annular, disk, and amorphous groups [12]. In the *Catalogue of Galactic Planetary Nebulae* published in 1967, Perek and Kohoutek classified PNe into stellar, disk, irregular, ring, or anomalous [8]. In the CCD era, Stanghellini et al. classified PNe into elliptical, bipolar, point-symmetric, irregular, and stellar groups [119]. Parker et al. classified the 903 new PNe in the *Macquarie/AAO/Strasbourg Hα Planetary Nebula Catalogue* (*MASH*) into six classes: round, elliptical, bipolar, irregular, asymmetric, and star-like [120]. The most basic classification is that of Balick, who groups PNe into round, elliptical, and butterfly classes [68]. Instead of a shell structure, bipolar PNe have a ring-like torus and a pair of lobes.

Morphological classifications of PNe are based on 2D images of PNe in the sky and are subject to the following uncertainties:

1. Images are sensitivity dependent, and a deeper image may reveal previously unseen faint features of different shapes;
2. Images taken with narrow-band filters may show different structures, as ionic species are distributed differently based on the ionization structure of the nebulae;
3. PNe are 3D objects and their apparent 2D morphology on the sky is dependent on the viewing angle;
4. The morphological classification may change when the nebulae are observed with a larger field of view;
5. Classifications are based on optical images and may not represent the distribution of the neutral (molecules and dust) component. The optical light distribution may not present an accurate picture of the actual matter distribution in the nebulae. Where we see light, does not mean that is where most of the matter is.

One question that has been debated by researchers for years is the fraction of bipolar nebulae in the PN population. When observed by CCD detectors, the well-known Messier object M76 (NGC 650-1) turns out to be bipolar, whereas earlier photographic plate images only showed the central torus of the bipolar nebula [58]. A bipolar nebula viewed near pole-on will be perceived to have a ring structure, due to the brightness of the torus. The inner haloes of the Ring Nebula NGC 6720 probably represent the projected images of the bipolar lobes [121] and the intrinsic structure of the Ring Nebula may actually resemble that of the bipolar Dumbbell Nebula NGC 6853 [122]. From a 3D model of kinematic data, the bipolar axis of the Ring Nebula is estimated to be inclined 10° relative to the line of sight [123]. The Helix Nebula NGC 7293 is now recognized as a bipolar nebula offset at 37°,

w.r.t. our line of sight [124,125]. The apparently spherical PN NGC 6369 is in fact a bipolar nebula when kinematic data are included in the modeling [126–128]. The 3D structures of several PNe have been constructed from imaging and spectro-kinematic observations [129].

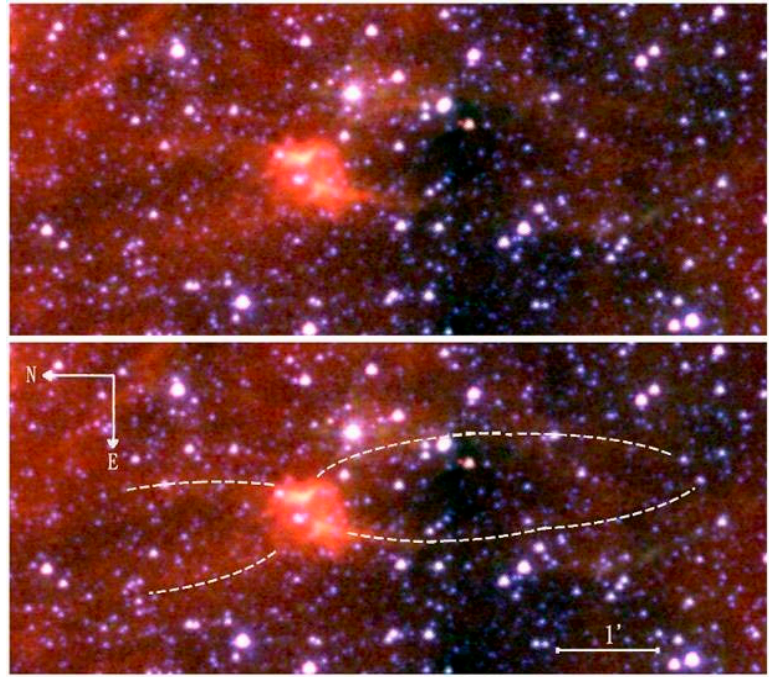

**Figure 4.** *Spitzer* IRAC color composite (3.6 μm in blue, 5.8 μm in green, and 8.0 μm in red) image of M 1-41. The bipolar structures extending over 3 arcmin are marked in the lower panel by dashed lines. Figure adapted from reference [130].

Since PN imaging is often limited by the field of view, wide field imaging may discover outer structures that may change the morphological classification of the nebula. A large spherical halo around NGC 7293 (Helix Nebula) was discovered in the *Wide-field Infrared Survey Explorer (WISE)* all-sky survey [131]. Large bipolar lobes extending over 100 arcsec were found outside of the main nebula of IPHAS PN-1 [132]. Figure 4 shows the color composite image of the *Spitzer Legacy Infrared Mid-Plane Survey Extraordinaire* (*GLIMPSE*) observation of PN M1-41. A large, extended structure can be seen well beyond the optical image of the nebula [130]. The *GLIMPSE* survey has also discovered many PNe that cannot be seen in the visible [133].

## 8. Multipolar Nebulae

High-dynamic range imaging of PNe reveals a new population of PNe with multiple bipolar lobes. The class of quadrupolar nebulae was identified by Manchado et al. [134], and multipolar nebulae were found by López et al. [135]. Most interestingly, multipolar lobes are found to be of approximately equal length within the nebula (e.g., in He 2–47, M1–37, He 2–113) [136]. Three examples of multiple nebulae are shown in Figure 5.

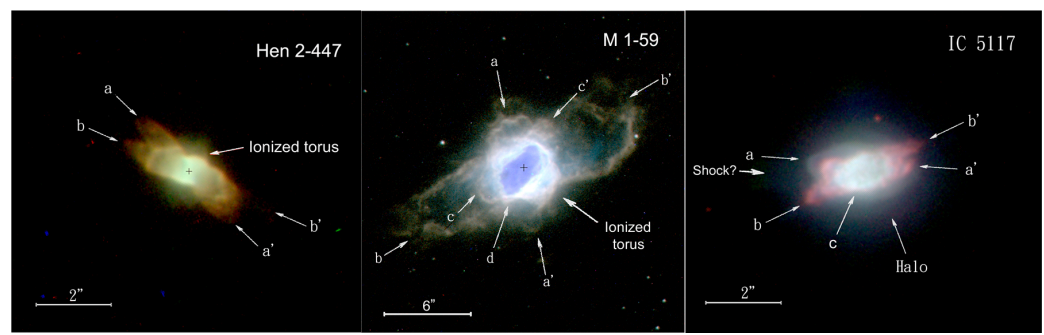

**Figure 5.** *HST* images of three multipolar nebulae. The pairs of lobes are labeled as (a, a'; b, b'; c, c' and d, d'). Figure adapted from reference [137].

If asymmetric PNe descend from AGB stars, which usually have spherically symmetrical stellar winds, then when did the shaping begin? The existence of multipolar nebulae suggests that the later-developed fast wind may be anisotropic, or highly collimated in certain directions. This has led to ideas on precessing jets as shaping agents in PN evolution, as in the model of bipolar, rotating episodic jets [138]. The near-equal sizes of the multipolar lobes suggests that these fast outflows are formed simultaneously or episodically [139].

Are the diverse morphological types of PNe the result of a single, unified 3D model? This idea was first explored by Khromov and Kohoutek using an open-ended toroid [140]. An ellipsoidal shell inclined w.r.t. the line of sight, with ionization penetrating to different depths, could also create different morphologies of PNe [141,142]. Recent studies show that when one takes into account the effects of orientation and sensitivity, a simple model of three pairs of identical lobes can reproduce many observed morphological features (such as tori, filaments, knots, and ansae) of PNe [143]. To obtain simultaneous spatial and kinematic information, integral field imaging spectroscopy can be used to construct a model of the true 3D structure of PNe.

## 9. The Role of Invisible Matter in Planetary Nebulae

From the *IRAS* and *Infrared Space Observatory* (*ISO*) observations of PNe, we have learnt that a significant fraction of energy emitted by PNe arises from the solid-state (dust) component. Figure 6 shows the spectral energy distribution of NGC 7027. The visible region is dominated by atomic emission lines and bound–free emissions from the ionized region of the nebula. The radio component is due to free–free emissions from the ionized nebula. In the wavelength region between 3 μm and 1 mm, thermal emission from the solid-state component dominates. A self-consistent model of NGC 7027 suggests that most of the mass of the nebula is in the neutral region [144].

Since the first detection of the CO molecule in NGC 7027 [145], we have come to realize that a significant fraction of the total mass of PNe can be in molecular form [146]. The extent of molecular distribution in PNe can be mapped by millimeter and submillimeter interferometers, such as the *Submillimeter Array* (*SMA*) [147,148] and *Atacama Large Millimeter/submillimeter Array* (*ALMA*) [149–151]. The cold dust component in PNe can be mapped by far-infrared imaging, and initial efforts in mid-infrared imaging have been made in ground-based observations [152–154], with *Spitzer* [155,156], *Herschel* [157,158], *AKARI* [159], and *SOFIA* [160,161], but the true extent of cool dust can only be revealed at far-infrared wavelengths with high sensitivity.

Our morphological classification of PNe is guided by optical images. Visible light from PNe originates from the ionized region through recombination lines of H and He and collisionally excited lines of metals [162]. These lines are bright and easy to see, but the total extent of the mass content of PNe must take into account the contribution from the neutral gas component.

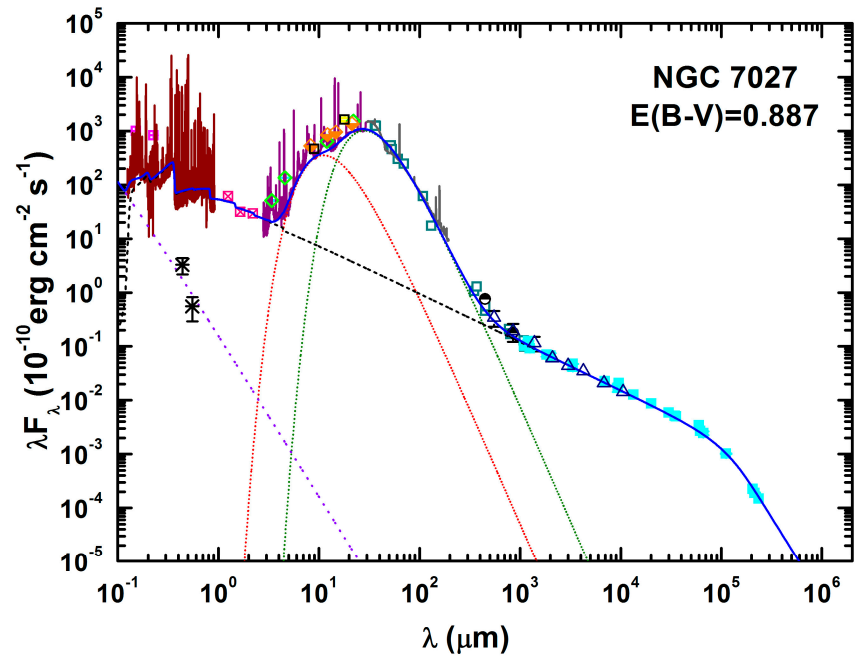

**Figure 6.** Spectral energy distribution of the planetary nebula NGC 7027 from wavelength 0.1 μm to 1 m. Various symbols represent photometric measurements and the lines are spectroscopic observations. The red and green curves are approximate blackbody fits to the dust continuum. The dashed line covering the entire spectrum is model bound–free and free–free emissions from the ionized nebula. The purple dashed line in the visible and UV region is an approximate fit to the central star continuum. The narrow features are atomic emission lines and some of the unidentified infrared emission (UIE) bands can be seen in the mid-infrared region. Molecular lines in the millimeter wave region are not plotted. The UV and optical photometry and spectroscopy have been corrected for circumstellar/interstellar reddening with E(B-V) = 0.887. Figure courtesy of C.H. Hsia.

Some optical nebulae show sharp boundaries, suggesting that the optical lobes are confined by an external medium. This is particularly evident in bipolar nebulae, where the tight waists of the nebulae are clearly confined (Figure 7). The optical lobes of bipolar nebulae are often interpreted as regions of massive ejections, but they could also represent low-density cavities carved out of the neutral circumstellar envelope by a collimated fast wind [163]. In this case, most of the mass of the PNe are in the unseen (dark) regions, outside of the optical lobes. Figure 8 shows three examples of optical lobes with well-defined boundaries, which suggests that the lobes are bound by unseen materials.

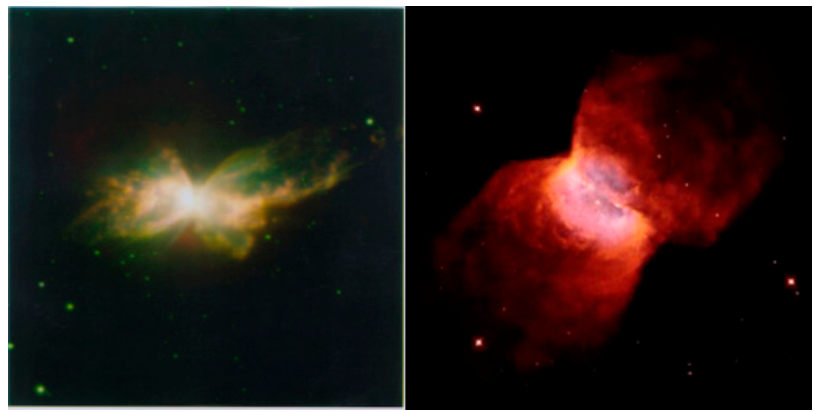

**Figure 7.** The well-known bipolar nebulae NGC 6302 (**left**) and NGC 2346 (**right**) show tight waists, which clearly suggests that they are confined by external media. Photo credits: NGC 6302 (Trung Hua), NGC 2346 (Hubble Heritage Team).

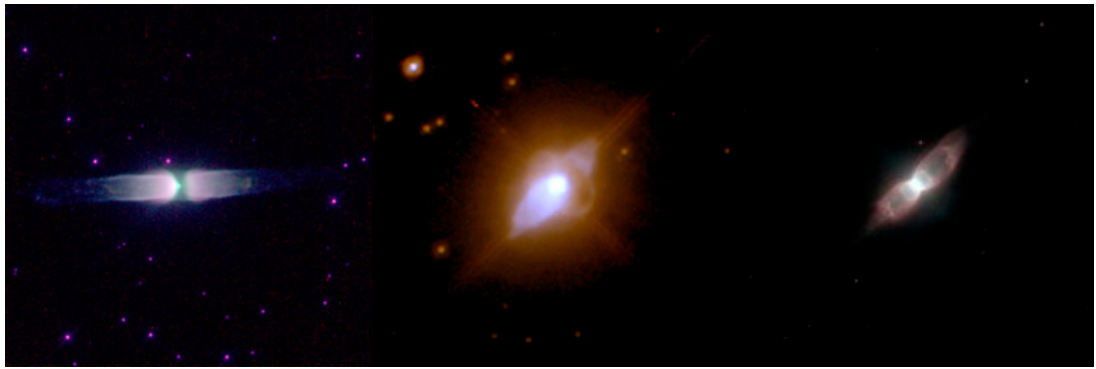

**Figure 8.** *HST* images of Hen 3-401 [72] (**left**), IRAS 17106–3046 [164] (**center**), and Hen 2–320 [137]. The sharp boundaries of the nebulae suggest that they are confined by external matter.

Other examples of bipolar nebulae, where the bipolar lobes could represent cavities bounded by this extensive circumstellar dust component, include M1−92 [165,166] and AFGL 618 [167,168]. In these objects, the optical images of the bipolar lobes can be compared with the molecular distribution mapped by millimeter wave interferometric observations. When viewed from different lines of sight, these objects could have very different apparent morphologies [169].

Since optical appearance is dependent on photoionization, the morphology of PNe will change with time as the central star becomes hotter. The magnitude and speed of the fast wind from the central star will also increase, leading to an eventual breakthrough of the ionization front. The low-surface brightness spherically shaped PNe found in deep H$\alpha$ surveys [120] may represent examples of highly evolved, density-bounded PNe.

## 10. Recent Progress and Still Unsolved Problems in PN Research

Progress in PN research has benefited greatly from technological advances, such as CCD cameras, infrared, millimeter wave, and X-ray observational instruments. Space-based observatories, such as *IRAS*, *ISO*, and *HST*, were responsible for many of the advances in the field. The recent deployment of *ALMA*, *Gaia*, and the *James Webb Space Telescope* (*JWST*) has also demonstrated their potential to bring about new discoveries.

Distances to PNe have been a long-standing problem, as statistical distances are often not accurate. The gravity distances based on model atmospheres of the central stars of PNe have provided good distances to some individual PNe [170]. Some success was also achieved with expansion distances based on multi-epoch *HST* observations [171]. From the results of the *Gaia* astrometric mission, reliable trigonometric distances to a large sample of PNe are finally available [172]. This leads to better determination of luminosities of PNe and their positions on the H-R diagram, according to which evolutionary tracks can be compared [173]. Figure 9 shows the distribution of 87 PNe in the H-R diagram. The distribution is covered by evolutionary tracks, with central star masses between 0.53 to 0.71 $M_\odot$.

Because of AGB mass loss, there is a large difference between the initial mass of stars and their final mass at the white dwarf stage. The determination of this initial–final mass relationship (IFMR) will provide us with answers on which stars will become PNe. For example, whether the Sun will become a PN is very much relevant to us. From studies on white dwarfs in open clusters, Volker Weidemann found an IFMR which suggests a much higher rate of mass loss in regard to the AGB than commonly believed at the time [174]. The recent availability of wide-field imagers and spectrographs on large telescopes has led to a much larger sample of white dwarfs and a more accurate determination of the IFMR [175].

Although many of the complex morphological features of PNe can be understood with the interacting winds framework, a complete understanding of the bipolar and multipolar structures of PNe remains a challenge. We know that the circumstellar envelope is transformed from a spherically symmetrical AGB wind into asymmetric shapes of PNe during the PPNe phase, but the exact mechanism of shaping is not known. The influence of a binary companion is commonly suggested, perhaps supplemented in some cases by a magnetic field [176]. Multipolar nebulae are probably created by a collimated wind, which may be variable in both time and direction. What physical mechanisms are driving these collimated outflows remain topics for further research.

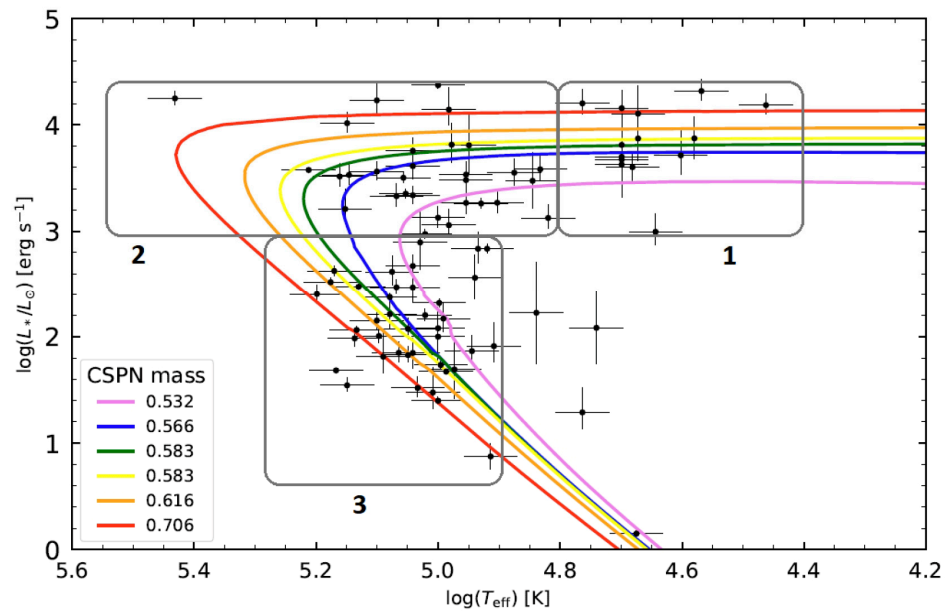

**Figure 9.** Distribution of planetary nebula central stars in the H-R diagram. The six evolutionary tracks are from Miller-Bertolami [32]. The corresponding central star masses of the tracks are marked in the legend. The total number of PNe plotted is 87. The three regions correspond to log ($L/L_\odot$) > 3.0 and log ($T_{eff}$) < 4.8 (region 1), log ($L/L_\odot$) > 3.0 and log ($T_{eff}$) > 4.8 (region 2), and log($L/L_\odot$) < 3.0 and log($T_{eff}$) > 4.9 (region 3). Figure adapted from reference [173].

Although PN research, in part, initiated the discipline of astrophysics in the early 20th century, PNe are now also recognized as a valuable laboratory for astrochemistry. PNe are prolific chemical factories. From the heavy elements synthesized during the AGB stage, molecules and solids are formed in the circumstellar nebula under a very low density and over very short time scales [177]. Rotational transitions of ~80 molecules have been detected with millimeter and submillimeter wave spectroscopic observations of AGB stars, proto-planetary nebulae, and PNe [178,179]. Spectral line surveys, with ground-based *ISO* and *Hershel* telescopes, have identified diverse classes of molecules, including inorganics, organics, radicals, chains, and rings. Molecules as large as $C_{60}$ and complex organics are synthesized in PNe and ejected into the interstellar medium, chemically enriching the Galaxy [180]. The unidentified infrared emission (UIE) bands, commonly observed in PNe, are likely to be vibrational bands of complex organics, but the exact chemical structure of their carrier has not been identified [181]. Although UIE bands are seen in H II regions, reflection nebulae, and in the diffuse interstellar medium, PNe and novae are the only objects to have been directly observed to show the UIE bands emerge in situ. Since UIE bands have been observed in galaxies as early as 1.5 billion years after the Big Bang [182], PNe could play a major role in organic synthesis in the Universe.

## 11. Conclusions

PNe are among the brightest extended objects in the sky. Modern optical spectroscopy can detect thousands of emission lines from the nebulae, allowing the determination of the abundance of many different chemical elements. Their brightness also makes it possible for very-high-dynamic range imaging and the simultaneous revelation of bright and faint features. Their high intrinsic brightness in narrow lines allow PNe to be detected at very large distances. Wide-field narrow-band surveys allow the detection of PNe in haloes of galaxies as far as the Virgo Cluster [183]. Extragalactic PNe have been used as distance indicators [184], for the determination of the Hubble constant, as estimates for baryonic mass, and to trace dark matter [185,186].

PNe also shines in all wavelengths, from radio to X-ray. This broad spectral coverage gives us a complete picture of the energetics of the objects, allowing PNe to serve as a laboratory for physical and chemical processes. A comprehensive model of PNe must consider radiation processes in all wavelengths and the interaction between all states of matter from ions, neutral atoms, molecules, to solids. The observation of the interactions of supersonic outflows in PNe can serve as testing ground for dynamical processes in the interstellar medium [162]. Lessons learned in PNe research have been applied to Wolf–Rayet nebulae, young stellar objects, and active galactic nuclei.

A model of PN should consider not just the visible component in ionized gases, but also matter in neutral atomic, molecular, and solid forms not seen in optical images. Although our attention is naturally drawn to the bright optical lobes, the optical lobes of bipolar/multipolar PN are not reliable indicators of mass distribution, as they could represent low-mass cavities carved out by fast winds. A comprehensive model can only be constructed from a combination of high-dynamic range optical imaging, molecular-line mapping in millimeter and submillimeter wavelengths, and far-infrared imaging.

Infrared spectroscopic observations of AGB stars, PPNe, and PNe, can help understand the chemical pathway leading from simple diatomic molecules to $C_{60}$ to aromatic rings to complex organics in the circumstellar environment. It is remarkable that chemical synthesis can occur under such low-density ($<10^6$ cm$^{-3}$) conditions and over very short ($\sim 10^3$ yr) time scales. The enrichment of the interstellar medium and primordial solar nebulae may have implications for the organic inventory of the Solar System and, maybe, even the primordial Earth [187].

Although PNe are bright objects and have been extensively studied for over 200 years, our path to an understanding of their nature has not been easy or straightforward. From their origin in catalogs of nebulous objects, PNe became the laboratory for astrophysics after the application of quantum atomic physics into nebular spectroscopic observations. In spite of advances in our understanding of nebular physics, our understanding of PNe as a phase of stellar evolution came much later. Our recognition of PN as a post-AGB phenomenon only happened after 1970. Our present models of the evolution of the central star and the dynamical evolution of the nebulae are the result of multi-wavelength observations that began in the 1970s.

Paradigm shifts do happen in astronomy. However, to challenge our existing paradigm of PN evolution, one has to present a self-consistent, alternative explanation of the morphology, abundance, kinematics, and statistics, of the entire PN phenomenon. It is remarkable that PN research has remained active and resilient after 200 years and has survived the many changes in scientific fashion. Whether PN research has a vital future depends on whether the PN community can find new angles to look at old problems, and whether we create new problems to solve through the application of new observational techniques.

**Funding:** This research was funded by the Natural Sciences and Engineering Research Council of Canada.

**Acknowledgments:** I would like to thank B.J. Hrivnak for providing valuable comments on an earlier draft of this paper. This paper is in part based on a colloquium given at the Instituto de Astronomía, Ensenada, Mexico. I would also like to thank Alberto Lopez for his kind hospitality.

**Conflicts of Interest:** The author declares that there are no conflicts of interest.

**References**

1. Hubble, E.P. The source of luminosity in galactic nebulae. *Astrophys. J.* **1922**, *56*, 400–438. [CrossRef]
2. Menzel, D.H. The Planetary Nebulae. *Publ. Astron. Soc. Pac.* **1926**, *38*, 295. [CrossRef]

3. Zanstra, H. An Application of the Quantum Theory to the Luminosity of Diffuse Nebulae. *Astrophys. J.* **1927**, *65*, 50. [CrossRef]
4. Bowen, I.S. The origin of the nebular lines and the structure of the planetary nebulae. *Astrophys. J.* **1928**, *67*, 1. [CrossRef]
5. Abell, G.O. Properties of Some Old Planetary Nebulae. *Astrophys. J.* **1966**, *144*, 259. [CrossRef]
6. Minkowski, R. The Sub-System of Planetary Nebulae. *Publ. Astron. Soc. Pac.* **1964**, *76*, 197. [CrossRef]
7. Henize, K.G. Observations of Southern Planetary Nebulae. *Astrophys. J. Suppl.* **1967**, *14*, 125. [CrossRef]
8. Perek, L.; Kohoutek, L. *Catalogue of Galactic Planetary Nebulae*; Academia: Praha, Czech Republic, 1967.
9. Acker, A.; Marcout, J.; Ochsenbein, F.; Stenholm, B.; Tylenda, R.; Schohn, C. *The Strasbourg-ESO Catalogue of Galactic Planetary Nebulae. Parts I, II*; ESO Publications: Garching, Germany, 1992.
10. Parker, Q.A. Planetary nebulae and how to find them: A concise review. *Front. Astron. Space Sci.* **2022**, *9*, 895287. [CrossRef]
11. Aller, L.H. *Gaseous Nebulae*; John Wiley and Sons: Hoboken, NY, USA, 1956.
12. Curtis, H.D. The Planetary Nebulae. *Publ. Lick Obs.* **1918**, *13*, 55–74.
13. Shklovsky, I.S. The nature of planetary nebulae and their nuclei. *Astron. Zh.* **1956**, *33*, 315.
14. Abell, G.O.; Goldreich, P. On the Origin of Planetary Nebulae. *Publ. Astron. Soc. Pac.* **1966**, *78*, 232. [CrossRef]
15. Shklovskii, I.S. Once More on the Distances to Planetary Nebulae and the Evolution of Their Nuclei. *Astron. Zh.* **1957**, *34*, 403.
16. O'Dell, C.R. The Evolution of the Central Stars of Planetary Nebulae. *Astrophys. J.* **1963**, *138*, 67. [CrossRef]
17. Harman, R.F.; Seaton, M.J. The ionization structure of planetary nebulae, IV. Optical thickness of the nebulae and temperatures of the central stars. *Mon. Not. R. Astron. Soc.* **1966**, *132*, 15. [CrossRef]
18. Seaton, M.J. The ionization structure of planetary nebulae. V. Radii, luminosities and problems of evolution. *Mon. Not. R. Astron. Soc.* **1966**, *132*, 113. [CrossRef]
19. Roxburgh, I.W. Origin of Planetary Nebulae. *Nature* **1967**, *215*, 838. [CrossRef]
20. Planetary Nebulae. *Nature* **1967**, *215*, 1438–1439. [CrossRef]
21. Salpeter, E.E. Evolution of Central Stars of Planetary Nebulae Theory. In *IAU Symposium 34: Planetary Nebulae*; Kluwer: Dordrechet, The Netherlands, 1968; pp. 409–420.
22. Deinzer, W.; Hansen, C. On the Evolution of the Central Stars of Planetary Nebulae. *Astron. J. Suppl.* **1968**, *73*, 173.
23. Kutter, G.S.; Savedoff, M.P.; Schuerman, D.W. A Mechanism for the Production of Planetary Nebulae. *Astrophys. Space Sci.* **1969**, *3*, 182–197. [CrossRef]
24. Shaviv, G. Theory of evolution of the central star. In *Symposium-International Astronomical Union*; Cambridge University Press: Cambridge, UK, 1978; pp. 195–199.
25. Kwok, S. On the distance of planetary nebulae. *Astrophys. J.* **1985**, *290*, 568–577. [CrossRef]
26. Paczyński, B. Evolution of Single Stars. VI. Model Nuclei of Planetary Nebulae. *Acta Astron.* **1971**, *21*, 417.
27. Schoenberner, D. Asymptotic giant branch evolution with steady mass loss. *Astron. Astrophys.* **1979**, *79*, 108–114.
28. Schoenberner, D. Late stages of stellar evolution: Central stars of planetary nebulae. *Astron. Astrophys.* **1981**, *103*, 119–130.
29. Kovetz, A.; Harpaz, A. The central star of a planetary nebula. *Astron. Astrophys.* **1981**, *95*, 66–68.
30. Iben, I., Jr. On the frequency of planetary nebula nuclei powered by helium burningand on the frequency of white dwarfs with hydrogen-deficient atmospheres. *Astrophys. J.* **1984**, *277*, 333–354. [CrossRef]
31. Wood, P.R.; Faulkner, D.J. Hydrostatic Evolutionary Sequences for the Nuclei of Planetary Nebulae. *Astrophys. J.* **1986**, *307*, 659. [CrossRef]
32. Miller Bertolami, M.M. New models for the evolution of post-asymptotic giant branch stars and central stars of planetary nebulae. *Astron. Astrophys.* **2016**, *588*, A25. [CrossRef]
33. Stasińska, G.; Gorny, S.K.; Tylenda, R. On the mass distribution of planetary nebulae central stars. *Astron. Astrophys.* **1997**, *327*, 736–742.
34. Weidmann, W.A.; Mari, M.B.; Schmidt, E.O.; Gaspar, G.; Miller Bertolami, M.M.; Oio, G.A.; Gutiérrez-Soto, L.A.; Volpe, M.G.; Gamen, R.; Mast, D. Catalogue of the central stars of planetary nebulae. *Astron. Astrophys.* **2020**, *640*, A10. [CrossRef]
35. Lucy, L.B. Formation of Planetary Nebulae. *Astron. J.* **1967**, *72*, 813. [CrossRef]
36. Paczyński, B.; Ziółkowski, J. On the Origin of Planetary Nebulae and Mira Variables. *Acta Astron.* **1968**, *18*, 255.
37. Kutter, G.S.; Sparks, W.M. Studies of Hydrodynamic Events in Stellar Evolution. III. Ejection of Planetary Nebulae. *Astrophys. J.* **1974**, *192*, 447–456. [CrossRef]
38. Wood, P.R. Models of Asymptotic-Giant Stars. *Astrophys. J.* **1974**, *190*, 609–630. [CrossRef]
39. Tuchman, Y.; Sack, N.; Barkat, Z. Miras and planetary nebula formation. *Astrophys. J.* **1979**, *234*, 217–227. [CrossRef]
40. Smith, R.L.; Rose, W.K. Relaxation Oscillations in the Envelopes of Luminous Red Giants. *Astrophys. J.* **1972**, *176*, 395. [CrossRef]
41. Faulkner, D.J. The Ejection of Planetary-Nebula Shells. *Astrophys. J.* **1970**, *162*, 513. [CrossRef]
42. Finzi, A.; Wolf, R.A. Ejection of Mass by Radiation Pressure in Planetary Nebulae. *Astron. Astrophys.* **1971**, *11*, 418.
43. Finzi, A.; Finzi, R.; Shaviv, G. The creation of planetary nebulae. *Astron. Astrophys.* **1974**, *37*, 325–334.
44. Harm, R.; Schwarzschild, M. Transition from a Red Giant to a Blue Nucleus after Ejection of a Planetary Nebula. *Astrophys. J.* **1975**, *200*, 324–329. [CrossRef]
45. Trimble, V.; Sackmann, I.-J. Ejection of planetary nebulae by helium shell flashes and the planetary distance scale. *Mon. Not. R. Astron. Soc.* **1978**, *182*, 97–102. [CrossRef]
46. Mathews, W.G. Model Planetary Nebulae. *Astrophys. J.* **1966**, *143*, 173. [CrossRef]
47. Capriotti, E.R. The structure and evolution of planetary nebulae. *Astrophys. J.* **1973**, *179*, 495. [CrossRef]

48. Ferch, R.L.; Salpeter, E.E. Models of planetary nebulae with dust. *Astrophys. J.* **1975**, *202*, 195. [CrossRef]
49. Wentzel, D.G. Dynamics of envelopes of planetary nebulae. *Astrophys. J.* **1976**, *204*, 452–460. [CrossRef]
50. Kwok, S.; Purton, C.R.; Fitzgerald, P.M. On the origin of planetary nebulae. *Astrophys. J. Lett.* **1978**, *219*, L125–L127. [CrossRef]
51. Kwok, S. From red giants to planetary nebulae. *Astrophys. J.* **1982**, *258*, 280–288. [CrossRef]
52. Kwok, S. *The Origin and Evolution of Planetary Nebulae*; Cambridge University Press: Cambridge, UK, 2000. [CrossRef]
53. Heap, S.R.; Boggess, A.; Holm, A.; Klinglesmith, D.A.; Sparks, W.; West, D.; Wu, C.C.; Boksenberg, A.; Willis, A.; Wilson, R.; et al. IUE observations of hot stars—HZ43, BD +75 deg 325, NGC 6826, SS Cygni, Eta Carinae. *Nature* **1978**, *275*, 385–388. [CrossRef]
54. Pottasch, S.R.; Baud, B.; Beintema, D.; Emerson, J.; Habing, H.J.; Harris, S.; Houck, J.; Jennings, R.; Marsden, P. IRAS measurements of planetary nebulae. *Astron. Astrophys.* **1984**, *138*, 10–18.
55. Zhang, C.Y.; Kwok, S. Spectral energy distribution of compact planetary nebulae. *Astron. Astrophys.* **1991**, *250*, 179.
56. Jewitt, D.C.; Danielson, G.E.; Kupferman, P.N. Halos around planetary nebulae. *Astrophys. J.* **1986**, *302*, 727–736. [CrossRef]
57. Chu, Y.-H.; Jacoby, G.H.; Arendt, R. Multiple-shell planetary nebulae. I—Morphologies and frequency of occurrence. *Astrophys. J. Suppl. Ser.* **1987**, *64*, 529–544. [CrossRef]
58. Balick, B.; Gonzalez, G.; Frank, A.; Jacoby, G. Stellar Wind Paleontology. II. Faint Halos and Historical Mass Ejection in Planetary Nebulae. *Astrophys. J.* **1992**, *392*, 582. [CrossRef]
59. Leahy, D.A.; Zhang, C.Y.; Kwok, S. Two-temperature X-ray emission from the planetary nebula NGC 7293. *Astrophys. J.* **1994**, *422*, 205–207. [CrossRef]
60. Conway, G.M.; Chu, Y.-H. X-ray emission from Planetary Nebulae. In *Proceedings of IAU Symposium 180: Planetary Nebulae*; Kluwer: Dordrecht, The Netherlands, 1997; pp. 214–215.
61. Kastner, J.H.; Montez, R., Jr.; Balick, B.; Frew, D.J.; Miszalski, B.; Sahai, R.; Blackman, E.; Chu, Y.-H.; De Marco, O.; Frank, A.; et al. The Chandra X-ray Survey of Planetary Nebulae (ChanPlaNS): Probing Binarity, Magnetic Fields, and Wind Collisions. *Astron. J.* **2012**, *144*, 58. [CrossRef]
62. Volk, K.; Kwok, S. Dynamical evolution of planetary nebulae. *Astron. Astrophys.* **1985**, *153*, 79–90.
63. Schmidt-Voigt, M.; Koeppen, J. Influence of stellar evolution on the evolution of planetary nebulae. I—Numerical method and hydrodynamical structures. II—Confrontation of models with observations. *Astron. Astrophys.* **1987**, *174*, 211–231.
64. Marten, H.; Schoenberner, D. On the dynamical evolution of planetary nebulae. *Astron. Astrophys.* **1991**, *248*, 590–598.
65. Perinotto, M.; Kifonidis, K.; Schonberner, D.; Marten, H. Hydrodynamical models of planetary nebulae and the problem of abundance determinations. *Astron. Astrophys.* **1998**, *332*, 1044–1054.
66. Marigo, P.; Girardi, L.; Groenewegen, M.A.T.; Weiss, A. Evolution of planetary nebulae—I. An improved synthetic model. *Astron. Astrophys.* **2001**, *378*, 958–985. [CrossRef]
67. Kahn, F.D.; West, K.A. Shapes of planetary nebulae. *Mon. Not. R. Astron. Soc.* **1985**, *212*, 837–850. [CrossRef]
68. Balick, B. The evolution of planetary nebulae. I—Structures, ionizations, and morphological sequences. *Astron. J.* **1987**, *94*, 671–678. [CrossRef]
69. Frank, A.; Mellema, G. A Radiation-Gasdynamical Method For Numerical Simulations Of Ionized Nebulae—Radiation-Gasdynamics Of PNE-I. *Astron. Astrophys.* **1994**, *289*, 937–945.
70. Mellema, G.; Frank, A. Radiation gasdynamics of planetary nebulae—V. Hot bubble and slow wind dynamics. *Mon. Not. R. Astron. Soc.* **1995**, *273*, 401–410. [CrossRef]
71. Mellema, G. The formation of bipolar planetary nebulae. *Astron. Astrophys.* **1997**, *321*, L29–L32. [CrossRef]
72. Kwok, S. *Cosmic Butterflies*; Cambridge University Press: Cambridge, UK, 2001; p. 190.
73. Corradi, R.L.M.; Schönberner, D.; Steffen, M.; Perinotto, M. Ionized haloes in planetary nebulae: New discoveries, literature compilation and basic statistical properties. *Mon. Not. R. Astron. Soc.* **2003**, *340*, 417–446. [CrossRef]
74. Frank, A.; Balick, B.; Riley, J. Stellar Wind Paleontology—Shells And Halos Of Planetary-Nebulae. *Astron. J.* **1990**, *100*, 1903–1914. [CrossRef]
75. Öttl, S.; Kimeswenger, S.; Zijlstra, A.A. Ionization structure of multiple-shell planetary nebulae. I. NGC 2438. *Astron. Astrophys.* **2014**, *565*, A87. [CrossRef]
76. Steffen, M.; Szczerba, R.; Schoenberner, D. Hydrodynamical models and synthetic spectra of circumstellar dust shells around AGB stars. II. Time-dependent simulations. *Astron. Astrophys.* **1998**, *337*, 149–177.
77. Corradi, R.L.M.; Schönberner, D.; Steffen, M.; Perinotto, M. A hydrodynamical study of multiple-shell planetaries. I. NGC 2438. *Astron. Astrophys.* **2000**, *354*, 1071–1085.
78. Schönberner, D.; Jacob, R.; Lehmann, H.; Hildebrandt, G.; Steffen, M.; Zwanzig, A.; Sandin, C.; Corradi, R.L.M. A hydrodynamical study of multiple-shell planetary nebulae. III. Expansion properties and internal kinematics: Theory versus observation. *Astron. Nachr.* **2014**, *335*, 378–408. [CrossRef]
79. Volk, K.M.; Kwok, S. Evolution of Proto--planetary Nebulae. *Astrophys. J.* **1989**, *342*, 345. [CrossRef]
80. Kwok, S. An infrared sequence in the late stages of stellar evolution. *Mon. Not. R. Astron. Soc.* **1990**, *244*, 179–183.
81. Hrivnak, B.J.; Kwok, S.; Volk, K.M. A study of several F and G supergiant-like stars with infrared excesses as candidates for proto-planetary nebulae. *Astrophys. J.* **1989**, *346*, 265–276. [CrossRef]
82. Kwok, S. Proto-planetary nebulae. *Annu. Rev. Astron. Astrophys.* **1993**, *31*, 63–92. [CrossRef]
83. Kwok, S.; Su, K.Y.L.; Hrivnak, B.J. Hubble Space Telescope V-Band Imaging of the Bipolar Proto-Planetary Nebula IRAS 17150-3224. *Astrophys. J. Lett.* **1998**, *501*, L117. [CrossRef]

84. Su, K.Y.L.; Volk, K.; Kwok, S.; Hrivnak, B.J. Hubble Space Telescope Imaging of IRAS 17441-2411: A Case Study of a Bipolar Nebula with a Circumstellar Disk. *Astrophys. J.* **1998**, *508*, 744–751. [CrossRef]
85. Hrivnak, B.J.; Kwok, S.; Su, K.Y.L. The Discovery of Two New Bipolar Proto-Planetary Nebulae: IRAS 16594-4656 and IRAS 17245-3951. *Astrophys. J.* **1999**, *524*, 849–856. [CrossRef]
86. Szczerba, R.; Siódmiak, N.; Stasińska, G.; Borkowski, J. An evolutionary catalogue of galactic post-AGB and related objects. *Astron. Astrophys.* **2007**, *469*, 799–806. [CrossRef]
87. Ueta, T.; Meixner, M.; Bobrowsky, M. A Hubble Space Telescope Snapshot Survey of Proto-Planetary Nebula Candidates: Two Types of Axisymmetric Reflection Nebulosities. *Astrophys. J.* **2000**, *528*, 861–884. [CrossRef]
88. Sahai, R.; Morris, M.; Sánchez Contreras, C.; Claussen, M. Preplanetary Nebulae: A Hubble Space Telescope Imaging Survey and a New Morphological Classification System. *Astron. J.* **2007**, *134*, 2200–2225. [CrossRef]
89. Siódmiak, N.; Meixner, M.; Ueta, T.; Sugerman, B.E.K.; Van de Steene, G.C.; Szczerba, R. Hubble Space Telescope Snapshot Survey of Post-AGB Objects. *Astrophys. J.* **2008**, *677*, 382–400. [CrossRef]
90. Su, K.Y.L.; Hrivnak, B.J.; Kwok, S. High-Resolution Imaging of Proto-Planetary Nebulae: The Effects of Orientation. *Astron. J.* **2001**, *122*, 1525–1537. [CrossRef]
91. Kwok, S.; Su, K.Y.L.; Stoesz, J.A. Circumstellar Arcs in AGB and Post-AGB Stars. In Proceedings of the Astrophysics and Space Science Library, Boston, MA, USA, 1 August 2001; p. 115.
92. Crabtree, D.R.; Rogers, C. Circumstellar Envelopes Observed as Optical Haloes. In Proceedings of the European Southern Observatory Conference and Workshop Proceedings, La Serena, Chile, 1 January 1993; p. 255.
93. Hrivnak, B.J.; Kwok, S.; Su, K.Y.L. The Discovery of Circumstellar Arcs around Two Bipolar Proto-planetary Nebulae. *Astron. J.* **2001**, *121*, 2775–2780. [CrossRef]
94. Sahai, R.; Trauger, J.T.; Watson, A.M.; Stapelfeldt, K.R.; Hester, J.J.; Burrows, C.J.; Ballister, G.E.; Clarke, J.T.; Crisp, D.; Evans, R.W.; et al. Imaging of the EGG Nebula (CRL 2688) with WFPC2/HST: A History of AGB/Post--AGB Giant Branch Mass Loss. *Astrophys. J.* **1998**, *493*, 301–311. [CrossRef]
95. Kwok, S.; Su, K.Y.L. Discovery of Multiple Coaxial Rings in the Quadrupolar Planetary Nebula NGC 6881. *Astrophys. J.* **2005**, *635*, L49–L52. [CrossRef]
96. Kwok, S.; Hsia, C.H. Multiple Coaxial Rings in the Bipolar Nebula Hubble 12. *Astrophys. J.* **2007**, *660*, 341–345. [CrossRef]
97. Koning, N.; Kwok, S.; Steffen, W. Morphology of the Red Rectangle Proto-planetary Nebula. *Astrophys. J.* **2011**, *740*, 27. [CrossRef]
98. Kraft, R.P. Binary Stars among Cataclysmic Variables. III. Ten Old Novae. *Astrophys. J.* **1964**, *139*, 457. [CrossRef]
99. Warner, B. *Cataclysmic Variable Stars*; Cambridge University Press: Cambridge, UK, 2003.
100. Han, Z.; Podsiadlowski, P. Binary Evolutionary Models. In *Proceedings of IAU Symposium 252: The Art of Modeling Stars in the 21st Century*; Cambridge University Press: Cambridge, UK, 2008; pp. 349–357.
101. Tutukov, A.V.; Yungelson, L.R. A Model for the Population of Binary Stars in the Galaxy. *Astron. Rep.* **2002**, *46*, 667–683. [CrossRef]
102. Han, Z.-W.; Ge, H.-W.; Chen, X.-F.; Chen, H.-L. Binary Population Synthesis. *Res. Astron. Astrophys.* **2020**, *20*, 161. [CrossRef]
103. Gallagher, J.S., III; Code, A.D. Ultraviolet photometry from the Orbiting Astronomical Observatory. X. Nova FH Serpentis 1970. *Astrophys. J.* **1974**, *189*, 303–314. [CrossRef]
104. Paczynski, B. *Common Envelope Binaries In IAU Symposium 73: Structure and Evolution of Close Binary Systems*; Reidel: Dredrecht, The Netherlands, 1976; p. 75.
105. Webbink, R.F. The Formation of White Dwarfs in Close Binary Systems. In *IAU Colloq. 53: White Dwarfs and Variable Degenerate Stars*; Kluwer: Dordrecht, The Netherlands, 1979; p. 426.
106. van den Heuvel, E.P.J. Late Stages of Close Binary Systems. In *IAU Symposium 73: Structure and Evolution of Close Binary Systems*; Reidel: Dordrecht, The Netherlands, 1976; p. 35.
107. Iben, I., Jr.; Tutukov, A.V. On the evolution of close binaries with components of initial mass between 3 solar masses and 12 solar masses. *Astrophys. J. Suppl. Ser.* **1985**, *58*, 661–710. [CrossRef]
108. Iben, I., Jr.; Tutukov, A.V. Binary stars and planetary nebulae. In *Proceedings of IAU Symposium 131: Planetary Nebulae*; Kluwer: Dordrecht, The Netherlands, 1989; pp. 505–522.
109. de Kool, M. Common Envelope Evolution and Double Cores of Planetary Nebulae. *Astrophys. J.* **1990**, *358*, 189. [CrossRef]
110. Bond, H.E.; Livio, M. Morphologies of Planetary Nebulae Ejected by Close-Binary Nuclei. *Astrophys. J.* **1990**, *355*, 568. [CrossRef]
111. De Marco, O. The Origin and Shaping of Planetary Nebulae: Putting the Binary Hypothesis to the Test. *Publ. Astron. Soc. Pac.* **2009**, *121*, 316–342. [CrossRef]
112. Jones, D.; Boffin, H.M.J. Binary stars as the key to understanding planetary nebulae. *Nat. Astron.* **2017**, *1*, 0117. [CrossRef]
113. Kohoutek, L. New and misclassified planetary nebulae. In *IAU Symposium 131 Planetary Nebulae*; Kluwe: Dordrecht, The Netherlands, 1987; pp. 29–37.
114. Frew, D.J.; Parker, Q.A. Planetary Nebulae: Observational Properties, Mimics and Diagnostics. *Publ. Astron. Soc. Aust.* **2010**, *27*, 129–148. [CrossRef]
115. Allen, D.A. A catalogue of symbiotic stars. *Astron. Soc. Aust. Proc.* **1984**, *5*, 369–421. [CrossRef]
116. Allen, D. Allen, D. A Perspective on the Symbiotic Stars. In *IAU Colloq. 103: The Symbiotic Phenomenon*; Kluwer: Dordrecht, The Netherlands, 1988; p. 3.
117. Feast, M.W.; Catchpole, R.M.; Whitelock, P.A.; Carter, B.S.; Roberts, G. The infrared variability and nature of symbiotic stars. V. Seven more systems. *Mon. Not. R. Astron. Soc.* **1983**, *203*, 373–383. [CrossRef]

118. Paczynski, B.; Rudak, B. Symbiotic Stars—Evolutionary Considerations. *Astron. Astrophys.* **1980**, *82*, 349–351.
119. Stanghellini, L.; Corradi, R.L.M.; Schwarz, H.E. The correlations between planetary nebula morphology and central star evolution. *Astron. Astrophys.* **1993**, *279*, 521–528.
120. Parker, Q.A.; Acker, A.; Frew, D.J.; Hartley, M.; Peyaud, A.E.J.; Ochsenbein, F.; Phillipps, S.; Russeil, D.; Beaulieu, S.F.; Cohen, M.; et al. The Macquarie/AAO/Strasbourg Hα Planetary Nebula Catalogue: MASH. *Mon. Not. R. Astron. Soc.* **2006**, *373*, 79–94. [CrossRef]
121. Bryce, M.; Balick, B.; Meaburn, J. Investigating the Haloes of Planetary Nebulae—Part Four—NGC6720 the Ring Nebula. *Mon. Not. R. Astron. Soc.* **1994**, *266*, 721. [CrossRef]
122. Kwok, S.; Chong, S.-N.; Koning, N.; Hua, T.; Yan, C.-H. The True Shapes of the Dumbbell and the Ring. *Astrophys. J.* **2008**, *689*, 219–224. [CrossRef]
123. Steffen, W.; López, J.A.; Koning, N.; Kwok, S.; Riesgo, H.; Richer, M.G.; Morisset, C. The 3D structure of the Ring Nebula. In Proceedings of the Asymmetrical Planetary Nebulae IV, La Palma, Spain, 1 June 2007; p. 38.
124. Meaburn, J.; Clayton, C.A.; Bryce, M.; Walsh, J.R.; Holloway, A.J.; Steffen, W. The nature of the cometary knots in the Helix planetary nebula (NGC7293). *Mon. Not. R. Astron. Soc.* **1998**, *294*, 201–223. [CrossRef]
125. Meaburn, J.; Boumis, P.; López, J.A.; Harman, D.J.; Bryce, M.; Redman, M.P.; Mavromatakis, F. The creation of the Helix planetary nebula (NGC 7293) by multiple events. *Mon. Not. R. Astron. Soc.* **2005**, *360*, 963–973. [CrossRef]
126. Monteiro, H.; Schwarz, H.E.; Gruenwald, R.; Heathcote, S. Three-Dimensional Photoionization Structure and Distances of Planetary Nebulae. I. NGC 6369. *Astrophys. J.* **2004**, *609*, 194–202. [CrossRef]
127. Steffen, W.; López, J.A. Morpho-Kinematic Modeling of Gaseous Nebulae with SHAPE. *Rev. Mexicana Astron. Astros.* **2006**, *42*, 99–105. [CrossRef]
128. Ramos-Larios, G.; Guerrero, M.A.; Vázquez, R.; Phillips, J.P. Optical and infrared imaging and spectroscopy of the multiple-shell planetary nebula NGC 6369. *Mon. Not. R. Astron. Soc.* **2012**, *420*, 1977–1989. [CrossRef]
129. Steffen, W. 3-D structures of planetary nebulae. *Proc. J. Phys. Conf. Ser.* **2016**, *728*, 032005. [CrossRef]
130. Zhang, Y.; Hsia, C.-H.; Kwok, S. Planetary Nebulae Detected in the Spitzer Space Telescope GLIMPSE 3D Legacy Survey. *Astrophys. J.* **2012**, *745*, 59. [CrossRef]
131. Zhang, Y.; Hsia, C.-H.; Kwok, S. Discovery of a Halo around the Helix Nebula NGC 7293 in the WISE All-sky Survey. *Astrophys. J.* **2012**, *755*, 53. [CrossRef]
132. Mampaso, A.; Corradi, R.L.M.; Viironen, K.; Leisy, P.; Greimel, R.; Drew, J.E.; Barlow, M.J.; Frew, D.J.; Irwin, J.; Morris, R.A.H.; et al. The "Príncipes de Asturias" nebula: A new quadrupolar planetary nebula from the IPHAS survey. *Astron. Astrophys.* **2006**, *458*, 203–212. [CrossRef]
133. Zhang, Y.; Kwok, S. Planetary Nebulae Detected in the Spitzer Space Telescope GLIMPSE II Legacy Survey. *Astrophys. J.* **2009**, *706*, 252–305. [CrossRef]
134. Manchado, A.; Stanghellini, L.; Guerrero, M.A. Quadrupolar Planetary Nebulae: A New Morphological Class. *Astrophys. J.* **1996**, *466*, L95. [CrossRef]
135. Lopez, J.A.; Meaburn, J.; Bryce, M.; Holloway, A.J. The Morphology and Kinematics of the Complex Polypolar Planetary Nebula NGC 2440. *Astrophys. J.* **1998**, *493*, 803–810. [CrossRef]
136. Sahai, R. The Starfish Twins: Two Young Planetary Nebulae with Extreme Multipolar Morphology. *Astrophys. J.* **2000**, *537*, L43–L47. [CrossRef]
137. Hsia, C.-H.; Chau, W.; Zhang, Y.; Kwok, S. Hubble Space Telescope Observations and Geometric Models of Compact Multipolar Planetary Nebulae. *Astrophys. J.* **2014**, *787*, 25. [CrossRef]
138. Lopez, J.A.; Vazquez, R.; Rodriguez, L.F. The Discovery of a Bipolar, Rotating, Episodic Jet (BRET) in the Planetary Nebula KjPn 8. *Astrophys. J.* **1995**, *455*, L63–L66. [CrossRef]
139. Sahai, R. Bipolar and Multipolar Jets in Protoplanetary and Planetary Nebulae. *Rev. Mex. Astron. Y Astrofis. Conf. Ser.* **2002**, *13*, 133–138.
140. Khromov, G.S.; Kohoutek, L. Morphological Study of Planetary Nebulae. In *IAU Symposium 34: Planetary Nebulae*; Reidel: Dordrecht, The Netherlands, 1968; pp. 227–235.
141. Masson, C.R. On The Structure Of Ionization-Bounded Planetary-Nebulae. *Astrophys. J.* **1990**, *348*, 580–587. [CrossRef]
142. Zhang, C.Y.; Kwok, S. A Morphological Study of Planetary Nebulae. *Astrophys. J. Suppl. Ser.* **1998**, *117*, 341. [CrossRef]
143. Chong, S.-N.; Kwok, S.; Imai, H.; Tafoya, D.; Chibueze, J. Multipolar Planetary Nebulae: Not as Geometrically Diversified as Thought. *Astrophys. J.* **2012**, *760*, 115. [CrossRef]
144. Volk, K.; Kwok, S. A Self-consistent Photoionization-Dust Continuum-Molecular Line Transfer Model of NGC 7027. *Astrophys. J.* **1997**, *477*, 722–731. [CrossRef]
145. Mufson, S.L.; Lyon, J.; Marionni, P.A. The detection of carbon monoxide emission in planetary nebulae. *Astrophys. J.* **1975**, *201*, L85–L89. [CrossRef]
146. Santander-García, M.; Jones, D.; Alcolea, J.; Bujarrabal, V.; Wesson, R. Lessons from the Ionised and Molecular Mass of Post-CE PNe. *Galaxies* **2022**, *10*, 26. [CrossRef]
147. Dinh-V-Trung; Bujarrabal, V.; Castro-Carrizo, A.; Lim, J.; Kwok, S. Massive Expanding Torus and Fast Outflow in Planetary Nebula NGC 6302. *Astrophys. J.* **2008**, *673*, 934–941. [CrossRef]

148. Wang, M.-Y.; Hasegawa, T.I.; Kwok, S. The Detection of a Molecular Bipolar Flow in the Multipolar Planetary Nebula NGC 2440. *Astrophys. J.* **2008**, *673*, 264–270. [CrossRef]
149. Santander-García, M.; Bujarrabal, V.; Alcolea, J.; Castro-Carrizo, A.; Sánchez Contreras, C.; Quintana-Lacaci, G.; Corradi, R.L.M.; Neri, R. ALMA high spatial resolution observations of the dense molecular region of NGC 6302. *Astron. Astrophys.* **2017**, *597*, A27. [CrossRef] [PubMed]
150. Moraga Baez, P.; Kastner, J.H.; Bublitz, J.; Alcolea, J.; Santander-Garcia, M.; Forveille, T.; Hily-Blant, P.; Balick, B.; Montez, R., Jr.; Gieser, C. ALMA Observations of Molecular Line Emission from High-excitation Bipolar Planetary Nebulae. *arXiv* **2024**, arXiv:2403.08961.
151. Andriantsaralaza, M.; Zijlstra, A.; Avison, A. CO in the C1 globule of the Helix nebula with ALMA. *Mon. Not. R. Astron. Soc.* **2020**, *491*, 758–772. [CrossRef]
152. Volk, K.; Hrivnak, B.J.; Su, K.Y.L.; Kwok, S. An Infrared Imaging Study of the Bipolar Proto-Planetary Nebula IRAS 16594-4656. *Astrophys. J.* **2006**, *651*, 294–300. [CrossRef]
153. Muthumariappan, C.; Kwok, S.; Volk, K. Subarcsecond Mid-Infrared Imaging of Dust in the Bipolar Nebula Hen 3-401. *Astrophys. J.* **2006**, *640*, 353–359. [CrossRef]
154. Lagadec, E.; Verhoelst, T.; Mékarnia, D.; Suáeez, O.; Zijlstra, A.A.; Bendjoya, P.; Szczerba, R.; Chesneau, O.; van Winckel, H.; Barlow, M.J.; et al. A mid-infrared imaging catalogue of post-asymptotic giant branch stars. *Mon. Not. R. Astron. Soc.* **2011**, *417*, 32–92. [CrossRef]
155. Su, K.Y.L.; Kelly, D.M.; Latter, W.B.; Misselt, K.A.; Frank, A.; Volk, K.; Engelbracht, C.W.; Gordon, K.D.; Hines, D.C.; Morrison, J.E.; et al. High spatial resolution mid- and far-infrared imaging study of NGC 2346. *Astrophys. J. Suppl. Ser.* **2004**, *154*, 302–308. [CrossRef]
156. Ueta, T. Spitzer MIPS Imaging of NGC 650: Probing the History of Mass Loss on the Asymptotic Giant Branch. *Astrophys. J.* **2006**, *650*, 228–236. [CrossRef]
157. van Hoof, P.A.M.; Van de Steene, G.C.; Exter, K.M.; Barlow, M.J.; Ueta, T.; Groenewegen, M.A.T.; Gear, W.K.; Gomez, H.L.; Hargrave, P.C.; Ivison, R.J.; et al. A Herschel study of NGC 650. *Astron. Astrophys.* **2013**, *560*, A7. [CrossRef]
158. Van de Steene, G.C.; van Hoof, P.A.M.; Exter, K.M.; Barlow, M.J.; Cernicharo, J.; Etxaluze, M.; Gear, W.K.; Goicoechea, J.R.; Gomez, H.L.; Groenewegen, M.A.T.; et al. Herschel imaging of the dust in the Helix nebula (NGC 7293). *Astron. Astrophys.* **2015**, *574*, A134. [CrossRef]
159. Ueta, T.; Torres, A.J.; Izumiura, H.; Yamamura, I.; Takita, S.; Tomasino, R.L. AKARI mission program: Excavating Mass Loss History in extended dust shells of Evolved Stars (MLHES). I. Far-IR photometry. *Publ. Astron. Soc. Jpn.* **2019**, *71*, 4. [CrossRef]
160. Werner, M.W.; Sahai, R.; Davis, J.; Livingston, J.; Lykou, F.; DE Buizer, J.; Morris, M.R.; Keller, L.; Adams, J.; Gull, G.; et al. Mid-infrared Imaging of the Bipolar Planetary Nebula M2-9 from SOFIA. *Astrophys. J.* **2014**, *780*, 156. [CrossRef]
161. Lau, R.M.; Werner, M.; Sahai, R.; Ressler, M.E. Evidence from SOFIA Imaging of Polycyclic Aromatic Hydrocarbon Formation along a Recent Outflow in NGC 7027. *Astrophys. J.* **2016**, *833*, 115. [CrossRef]
162. Kwok, S. *Physics and Chemistry of the Interstellar Medium*; University Science Books: San Francisco, CA, USA, 2007.
163. Kwok, S. Morphological Structures of Planetary Nebulae. *Publ. Astron. Soc. Aust.* **2010**, *27*, 174–179. [CrossRef]
164. Kwok, S.; Hrivnak, B.J.; Su, K.Y.L. Discovery of a Disk-collimated Bipolar Outflow in the Proto-Planetary Nebula IRAS 17106-3046. *Astrophys. J.* **2000**, *544*, L149–L152. [CrossRef]
165. Bujarrabal, V.; Alcolea, J.; Sahai, R.; Zamorano, J.; Zijlstra, A.A. The shock structure in the protoplanetary nebula M 1-92: Imaging of atomic and H-2 line emission. *Astron. Astrophys.* **1998**, *331*, 361–371.
166. Alcolea, J.; Neri, R.; Bujarrabal, V. Minkowski's footprint revisited. Planetary nebula formation from a single sudden event? *Astron. Astrophys.* **2007**, *468*, L41–L44. [CrossRef]
167. Sánchez Contreras, C.; Sahai, R.; Gil de Paz, A. Physical Structure of the Proto-Planetary Nebula CRL 618. I. Optical Long-Slit Spectroscopy and Imaging. *Astrophys. J.* **2002**, *578*, 269–289. [CrossRef]
168. Bujarrabal, V.; Alcolea, J.; Soria-Ruiz, R.; Planesas, P.; Teyssier, D.; Marston, A.P.; Cernicharo, J.; Decin, L.; Dominik, C.; Justtanont, K.; et al. Herschel/HIFI observations of high-J CO transitions in the protoplanetary nebula CRL 618. *Astron. Astrophys.* **2010**, *521*, L3. [CrossRef]
169. Koning, N.; Kwok, S.; Steffen, W. Post Asymptotic Giant Branch Bipolar Reflection Nebulae: Result of Dynamical Ejection or Selective Illumination? *Astrophys. J.* **2013**, *765*, 92. [CrossRef]
170. Mendez, R.H.; Kudritzki, R.P.; Herrero, A. On central star luminosities and optical thicknesses in planetary nebulae. *Astron. Astrophys.* **1992**, *260*, 329–340.
171. Schönberner, D.; Balick, B.; Jacob, R. Expansion patterns and parallaxes for planetary nebulae. *Astron. Astrophys.* **2018**, *609*, A126. [CrossRef]
172. Kimeswenger, S.; Barría, D. Planetary nebula distances in Gaia DR2. *Astron. Astrophys.* **2018**, *616*, L2. [CrossRef]
173. González-Santamaría, I.; Manteiga, M.; Manchado, A.; Ulla, A.; Dafonte, C. Gaia DR2 Distances to Planetary Nebulae. *Galaxies* **2020**, *8*, 29. [CrossRef]
174. Weidemann, V. Mass loss towards the white dwarf stage. *Astron. Astrophys.* **1977**, *59*, 411–418.
175. Cummings, J.D.; Kalirai, J.S.; Tremblay, P.-E.; Ramirez-Ruiz, E.; Choi, J. The White Dwarf Initial-Final Mass Relation for Progenitor Stars from 0.85 to 7.5 $M_\odot$. *Astrophys. J.* **2018**, *866*, 21. [CrossRef]

176. Decin, L.; Montargès, M.; Richards, A.M.S.; Gottlieb, C.A.; Homan, W.; McDonald, I.; El Mellah, I.; Danilovich, T.; Wallström, S.H.J.; Zijlstra, A.; et al. (Sub)stellar companions shape the winds of evolved stars. *Science* **2020**, *369*, 1497–1500. [CrossRef]
177. Kwok, S. The synthesis of organic and inorganic compounds in evolved stars. *Nature* **2004**, *430*, 985–991. [CrossRef]
178. Ziurys, L.M. The chemistry in circumstellar envelopes of evolved stars: Following the origin of the elements to the origin of life. *Proc. Natl. Acad. Sci. USA* **2006**, *103*, 12274–12279. [CrossRef]
179. Cernicharo, J.; Agúndez, M.; Guélin, M.; Bachiller, R. Spectral Line Surveys of Evolved Stars. In *IAU Symposium 280: The Molecular Universe*; Cambridge University Press: Cambridge, UK, 2011; pp. 237–248.
180. Kwok, S. Planetary Nebulae as Sources of Chemical Enrichment of the Galaxy. *Front. Astron. Space Sci.* **2022**, *9*, 893061. [CrossRef]
181. Kwok, S. The mystery of unidentified infrared emission bands. *Astrophys. Space Sci.* **2022**, *367*, 16. [CrossRef]
182. Spilker, J.S.; Phadke, K.A.; Aravena, M.; Archipley, M.; Bayliss, M.B.; Birkin, J.E.; Béthermin, M.; Burgoyne, J.; Cathey, J.; Chapman, S.C.; et al. Spatial variations in aromatic hydrocarbon emission in a dust-rich galaxy. *Nature* **2023**, *618*, 708–711. [CrossRef]
183. Hartke, J.; Arnaboldi, M.; Longobardi, A.; Gerhard, O.; Freeman, K.C.; Okamura, S.; Nakata, F. The halo of M 49 and its environment as traced by planetary nebulae populations. *Astron. Astrophys.* **2017**, *603*, A104. [CrossRef]
184. Jacoby, G.H.; Branch, D.; Ciardullo, R.; Davies, R.L.; Harris, W.E.; Pierce, M.J.; Pritchet, C.J.; Tonry, J.L.; Welch, D.L. A Critical Review of Selected Techniques for Measuring Extragalactic Distances. *Publ. Astron. Soc. Pac.* **1992**, *104*, 599. [CrossRef]
185. Meatheringham, S.J.; Dopita, M.A.; Ford, H.C.; Webster, B.L. The Kinematics of the Planetary Nebulae in the Large Magellanic Cloud. *Astrophys. J.* **1988**, *327*, 651–663. [CrossRef]
186. Coccato, L. Planetary nebulae as kinematic tracers of galaxy stellar halos. In *IAU Symposium 323: Planetary Nebulae: Multi-Wavelength Probes of Stellar and Galactic Evolution*; Cambridge University Press: Cambridge, UK, 2017; pp. 271–278.
187. Kwok, S. Delivery of Complex Organic Compounds from Evolved Stars to the Solar System. *Orig. Life Evol. Biosph.* **2011**, *41*, 497–502. [CrossRef]